# Peeping into the SU(2) Gauge Vacuum [*][†]

A. González-Arroyo [a] and P. Martínez

[a] Departamento de Física Teórica C-XI,
Universidad Autónoma de Madrid,
28049 Madrid, Spain.

We study thermalised configurations of SU(2) gauge fields by cooling. An analysis of the effect of cooling is presented and global and statistical information is extracted.

## 1. Introduction

It is highly desirable to have a qualitative understanding of the mechanism underlying the occurence of non-perturbative phenomena like Confinement, spontaneous chiral symetry breaking, the $\eta'$ mass generation, etc., and a consequent estimation of the relevant physical quantities: string tension, topological susceptibility, $<\bar{\psi}\psi>$, etc. Many authors have attempted this program, giving a description of the physical Yang-Mills vacuum, and identifying what they believe are the relevant degrees of freedom to understand all these phenomena. The lattice formulation has been able of determining most of these physical constants numerically out of the QCD lagrangian in fairly good agreement in most cases with the expectations. Hence, it seems that the lattice configurations which give rise to these results contain the essential information to discriminate between the different mechanisms and models. Some of these models are based in attributing a fundamental role to a given class of classical configurations, and it should be possible a priori to identify them and check that they actually do as expected.

In practice, the problem becomes very hard since lattice configurations are very noisy and it is hard to see any pattern in them. However, most authors agree that the noise is basically short ranged and does not affect long distance correlations, which lie in the heart of all non-perturbative behaviour. The cooling method was introduced precisely as a method to get rid of the short range fluctuations while keeping the long range ones unaltered, and hence is the instrument used by most authors to attempt this program. Our work complies to this scheme.

We present preliminary results based on the analysis of 56 configurations of SU(2) gauge fields. These configurations are generated by a heat bath method and Wilson's action at $\beta = 2.325$ and a $8^3 \times 64$ lattice and with a $\vec{m} = (1,1,1)$ twist. Our configurations are separated by at least 7500 Monte Carlo iterations and all our checks imply that they are fairly uncorrelated. The choice of parameters is based on our previous work [2] in which we studied SU(2) gauge fields for this twist value and with various $\beta$ values and lattice sizes. There we saw that at the present values results have small finite size effects.

## 2. The effect of cooling

Since we are going to use cooling, our first concern is to know to what extent does cooling affect the results which are obtained. For that purpose we have carried on several checks and tests by applying the cooling procedure to known classical configurations. We will actually be playing with various versions of cooling based of actions different from Wilson's one. Our family of coolings [1] depends on one parameter $\epsilon$ which is equal to 1 for Wilson's action and to 0 for a classically improved action ( See Ref. 1 for details). Negative values of $\epsilon$ lead to opposite sign contributions to the action and are hence called over-improved.

[*]Based on the talk given by A. González-Arroyo
[†]Partially financed by CICYT and EC-HCM Contract CHRX-CT92-0051

This additional freedom is particularly important when the family of classical configurations contains zero modes, since then O(a) corrections become leading. In particular, in Ref. 1 we saw that when considering sufficiently smooth lattice instantons, ordinary cooling causes the width to decrease, while for over-improved values the tendency is reversed.

Our first test involved putting an isolated instanton on the lattice and studying the way in which it evolved with the different coolings. We have started with a lattice instanton extracted from our Monte Carlo data, and cooled it with $\epsilon = 1$ until it became fairly small ( Thin Instanton). Then we reversed the sign of $\epsilon$ and cooled for 1000 steps (Fat Instanton). Next, we cooled back with +1 until the instanton eventually disappeared. For each step we computed the quantity $R = 6\pi^2/E^{max}$, where $E^{max}$ is the maximum of the space integrated energy. For a continuum instanton $R$ is equal to $\rho$. We also computed a series of improved versions of the topological charge, which for smooth fields reproduce the continuum formula to order $a^{2n}$. The formula is based on computing the Chromo-Electric and Magnetic field strengths by a clover average of square plaquettes of sizes 1, 2, ...n. and then computing the integral of the scalar product in space and color of both fields: $Q^{(n)}$. Fig. 1 displays the history profile, where the horizontal axis counts cooling steps and the kinks mark the points where $\epsilon$ was reversed. Notice that for large enough instantons the width varies linearly. By varying the value of $\epsilon$ we have found that the rate of growth or decrease depends linearly with $\epsilon$ although with a different proportionality for positive and negative $\epsilon$.

For small instantons the behaviour is more complicated. We have analised the result of cooling with various $\epsilon$ the Thin Instanton configuration. Actually all values larger that $-0.3$ lead to a destruction of the configuration in a few tens of cooling steps. We conclude with the following panorama: Positive values of $\epsilon$ lead to an eventual destruction of all instantons in a calculable number of steps depending on the original size. For negative $\epsilon$ values there is a minimal size below which they are unstable under cooling. Based

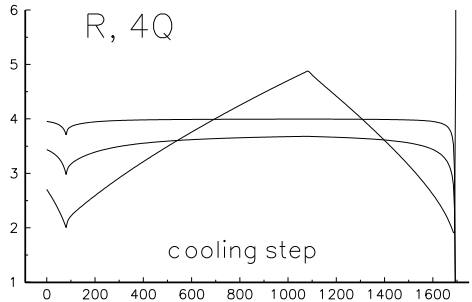

Figure 1. $R$ (thick line) and $4 Q^{(2)}$ ( top thin) and $Q^{(1)}$, as a function of cooling step for a lattice instanton configuration. The value of $\epsilon$ changed from 1 to -1 and back to 1.

on these results we have chosen $\epsilon = -0.3$ as the value to be used in the remaining.

It is also interesting to realize that due to the size variation, $Q^{(1)}$ will also vary with cooling inducing a variation of the topological susceptibility. It is recommended to use $Q^{(2)}$ which is very slightly $\rho$ dependent and close to 1 even up to very small sizes.

## 3. Global Quantities

We have applied up to 50 cooling steps with $\epsilon = -0.3$ on the set of 56 configurations, and analysed the dependence of several observables with the number of cooling steps. We have measured the topological susceptibility associated to each of the clover averaged charges:

$$\chi^{(n)} = \frac{(Q^{(n)})^2}{Volume} \qquad (1)$$

In Fig. 2 we show the results of $\chi^{(2)}$ as a function of cooling steps, compared to the results obtained with Wilson's action cooling. There is a sizable difference between both, which can be attributed to the destruction of instantons in the second case. Using $\chi^{(1)}$ instead of $\chi^{(2)}$ gives differences of the same magnitude. We suggest that the best choice would be to take $\epsilon = -0.3$ and cool for a very large number of times. This should produce a stable value of the susceptibility close to the one we found at 50 coolings. This value does not include the effect of instantons of sizes smaller than the Thin Instanton ( $\sim 2\,a$ ), which have been eliminated by the cooling procedure. One should then estimate the contribution of these small instantons. In principle, asymptotic freedom allows to determine the distribution of small enough instantons.

Now we turn to other observables: The electric flux $\vec{e}$ sector energies, measured by looking at the correlation of spatial Polyakov loops of winding equal to $\vec{e}$. We have measured the following quantities for all cooling steps up to 50

$$\Sigma_e(n_t) = -\frac{N_s}{e\,l s^2} ln(\frac{C_e(n_t)}{C_e(n_t-1)}) \qquad (2)$$

where $e \equiv |\vec{e}|$ and $C_e(n_t)$ is the lattice correlation function at distance $n_t$ for Polyakov loops of flux $\vec{e}$. The physical distance $l_s \equiv a(\beta) N_s$ is fixed by setting the infinite volume string tension to $\sigma = 5 fm.^{-2}$. In the infinite volume limit $\Sigma_e(n_t)$ should tend to $\sigma$ for large $n_t$ independently of $e$. Our results are indeed quite compatible for all three fluxes, confirming again that finite size effects are small. $\Sigma_e(n_t)$ grows with $n_t$ for small values, but the data seems to flatten out for larger ones. Indeed, the result for $n_t = 5$ is compatible with the larger distance ones, which have large errors. Taking the value of $\Sigma_e(5)$ as an estimate of the string tension, we could conclude that it decreases with cooling, being for 50 cooling steps at about $40 - 60\%$ of its uncooled value. Since the configurations at fifty coolings are very closely given by a bunch of self-dual (instantonic) lumps, it seems that these objects account for at least 50% of the value of the string tension. Of course these instantonic structures must be highly correlated in order to give rise to a finite string tension. In Ref.2 we gave an scenario in which this can occur. Some work is needed to clarify the reason for the slow decrease that is still observed as a function of the number of coolings.

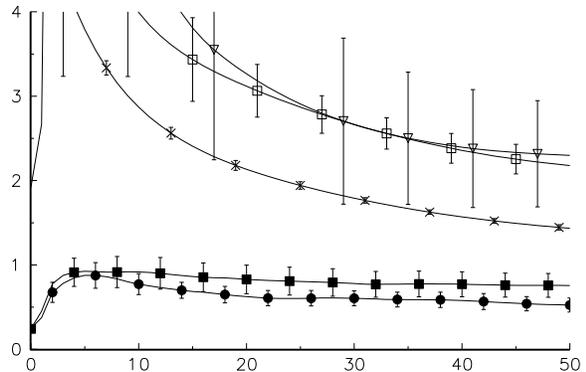

Figure 2. The topological susceptibility $\chi^{(2)}$ for $\epsilon = -0.3$ ( filled squares) and $\epsilon = 1$ (filled circles), and the quantities $\Sigma_1(n_t)$ for $n_t = 3, 5$ and 7 (crosses, squares and tringles respectively) as a function of cooling steps. All quantities are given in appropiate fermi units.

## 4. Statistical Distributions

We have performed a number of statistical analysis on our data sample, which cannot be presented in such short space. We found that the electric and magnetic field densities are highly correlated, but there is a slight excess of magnetic energy. Identifying energy peaks with instantons, we obtained that the distribution of widths is fairly flat in contrast with the results of Ref. 3. The distribution in space is also flat beyond $8\,a$, showing both attractive and repulsive region below.